\documentclass[12pt]{iopart}

\usepackage{iopams}  
\usepackage{graphicx}
\usepackage{dcolumn}
\usepackage{bm}
\begin{document}

\title[A Simple Model for Cavity Enhanced Slow Lights in VCSELs]{A Simple Model for Cavity Enhanced Slow Lights in Vertical Cavity Surface Emission Lasers}
\author{Chia-Sheng Chou and Ray-Kuang Lee}
\address{Institute of Photonics Technologies, National Tsing-Hua University, Hsinchu, 300 Taiwan}
\author{Peng-Chun Peng}
\address{Department of Applied Materials and Optoelectronic Engineering, National Chi Nan University, Nantou, 545 Taiwan}
\author{Hao-chung Kuo}
\address {Department of Photonics and Institute of Electro-Optical Engineering, National Chiao-Tung University, Hsinchu, 300 Taiwan}
\author{Gray Lin}
\address{Department of Electronics Engineering, National Chiao-Tung University, Hsinchu, 300 Taiwan}
\author{Hung-Ping Yang and Jim Y. Chi}
\address{Opto-Electronics and System Laboratory, Industrial Technology Research Institute, Hsinchu, 300 Taiwan}
\ead{rklee@ee.nthu.edu.tw}

\begin{abstract}
We develop a simple model for the slow lights in Vertical Cavity Surface Emission Lasers (VCSELs), with the combination of cavity and population pulsation effects.
The dependences of probe signal power, injection bias current and wavelength detuning for the group delays are demonstrated numerically and experimentally.
Up to $65$ ps group delays and up to 10 GHz modulation frequency can be achieved in the room temperature at the wavelength of $1.3$ $\mu$m.
The most significant feature of our VCSEL device is that the length of active region is only several $\mu$m long.
Based on the experimental parameters of quantum dot VCSEL structures, we show that the resonance effect of laser cavity plays a significant role to enhance the group delays.
\end{abstract} 

\vspace{2pc}
\submitto{\JOA\\Special Issue from Optical MEMS and Nanophotonics 2007}
\maketitle
\date{\today}

\section{Introduction}
Slow light is believed to be a critical foundation not only for basic scientific research but also for applications in optical communication, optical memories, signal processing, and phase-array antenna systems \cite{Boyd}.
Various systems have been demonstrated for slow lights, from electromagnetically induced transparency (EIT) \cite{Harris, Hau}, coherent population oscillations (CPO) \cite{Bigelow}, to stimulated Brillouin \cite{Brillouin} and Raman scatterings \cite{Raman}.
Unlike EIT in the cryogenic systems, slow light in semiconductor optoelectronic devices based on CPO is more promising due to its inherent compactness, direct electrical controllability, and room temperature operation.
CPO is the effect that the ground state population of the material will oscillate in time at the beat frequency of the two input waves.
This involves shining two lasers - a pump beam and a weaker probe beam - at the media.
The probe beam experiences reduced absorption over a narrow range of wavelengths under certain conditions.
The refractive index also increases rapidly in this {\it spectral hole}, which leads to a much reduced group velocity for the probe beam.

With state of art fabrication technologies, quantum well and quantum dot semiconductor optical amplifiers (SOAs) have been demonstrated as a flexible platform for studying slow light phenomenon based on CPO as well as its applications in room temperature \cite{Ku, SLC-apl06}.
For quantum dot (QD) can provide a better carrier confinement and offer reduced thermal ionization, semiconductor lasers with quantum dot gain media have been studied intensively to improve the laser characteristics.
Recently we demonstrate a tunable optical group delay in the monolithically single mode quantum dot Vertical Cavity Surface Emitting Laser (VCSEL) at $10$ GHz experimentally \cite{PCP-oe}.
Tunable slow light with optical group delay up to several tens of picoseconds can be achieved by adjusting the bias current and wavelength detuning.

The main difference between SOA and VCSEL devices is that the latter one has a cavity induced by two Bragg gratings.
Compared to SOA devices with the active region for the gain medium about several $mm$, instead the active region of VCSEL is typically only several $\mu$m long.
In this scenario, the common adopted population pulsation model of traveling waves induced dynamic carrier index grating \cite{Agrawal} can not directly applied to semiconductor lasers.
Followed by the two-wave model for the pump and probe beams in the presence of coherent population oscillation, in this work we develop a simple model for the slow lights in VCSELs with the combination of cavity effect and the rate equation for carrier undulation.
A simple formulation based on a Fabry-Perot filter with gain medium within \cite{Laurand06} is used to model real distributed Bragg reflectors in our VCSELs.
Experimental data of up to $65$ ps group delays and up to $10$ GHz modulation frequency operated in the room temperature at the wavelength of $1.3$ $\mu$m are in agreement with the proposed theoretical results.
Based on the experimental parameters of quantum dot VCSEL structures, we show that the resonance effect of laser
cavity plays a significant role to enhance the group delays. 
\begin{figure}
\includegraphics[width=8.0cm]{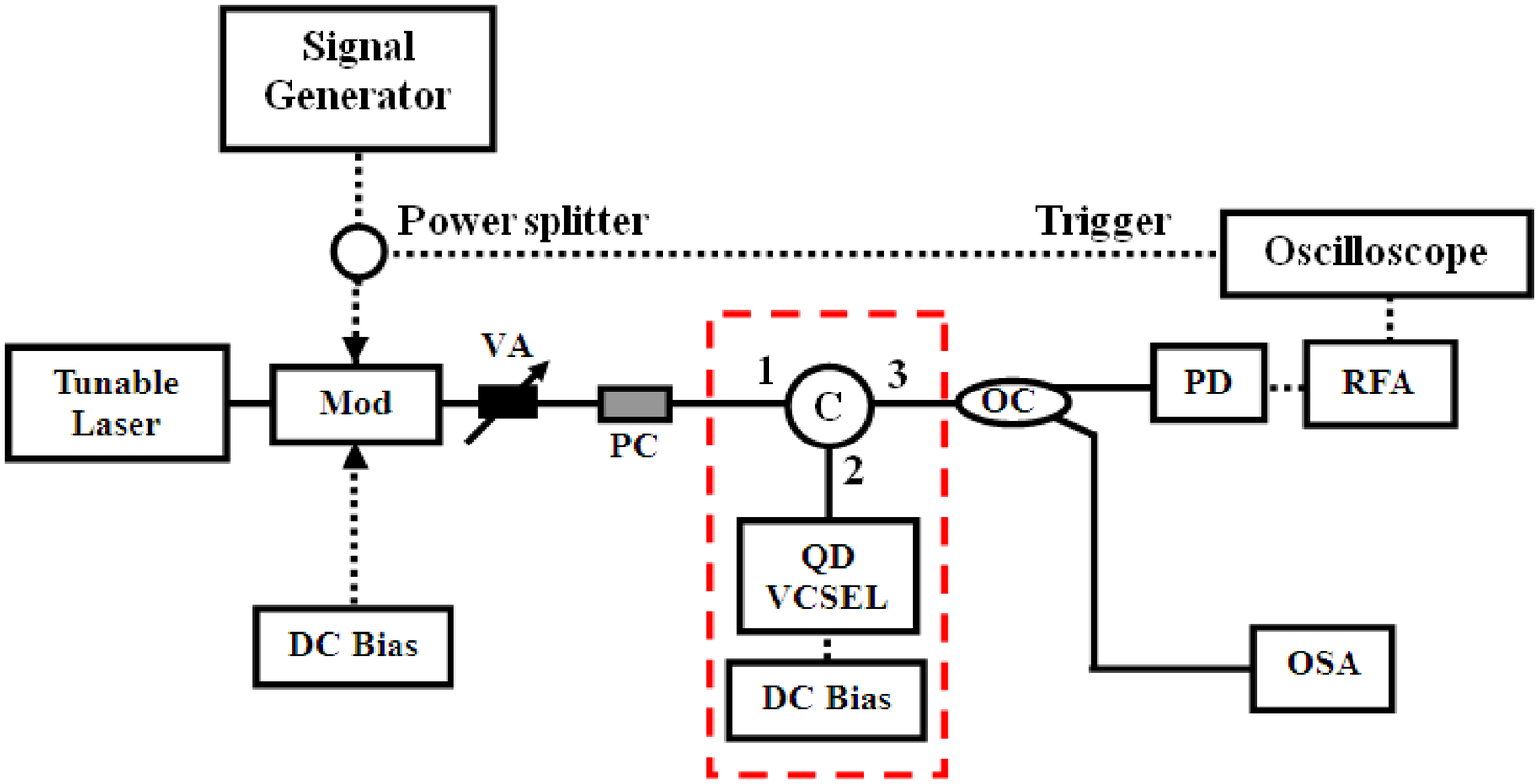}
\includegraphics[width=8.0cm]{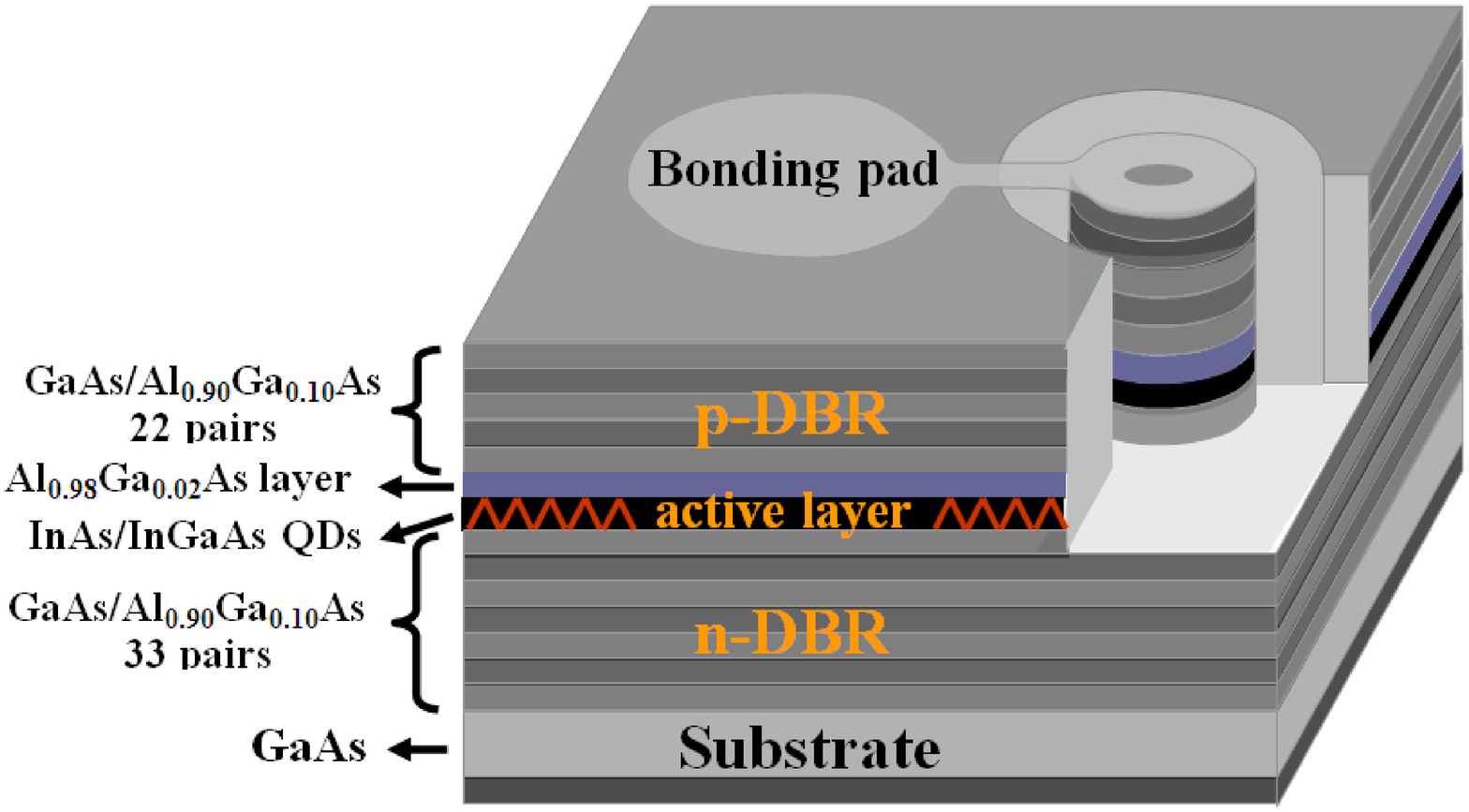}
\caption{(a) Experimental setup for measuring the optical group delays in VCSELs. (Mod: electro-optic modulator, VA: variable optical attenuator, C: optical circulator, OC: optical coupler, PC: polarization controller, RFA: RF amplifier, PD: photodetector, OSA: optical spectrum analyzer) (b) Schematic diagram of our quantum dot VCSEL.}
\label{Fig:F1}
\end{figure}
\section{Fabrication and Measurement of the slow light in VCSELs}
The experiment setup for the slow-light in VCSELs is illustrated in Fig. \ref{Fig:F1}(a).
The key component in our experiment is a monolithically single-mode GaAs based QD VCSEL, grown by molecular beam epitaxy (MBE) with fully doped n- and p-doped AlGaAs distributed Bragg reflectors (DBRs), as show in Fig. \ref{Fig:F1}(b).
The characteristics of our QD VCSEL have been described in our previous works \cite{PCP-long, PCP-el}.

\begin{figure}
\includegraphics[width=8.0cm]{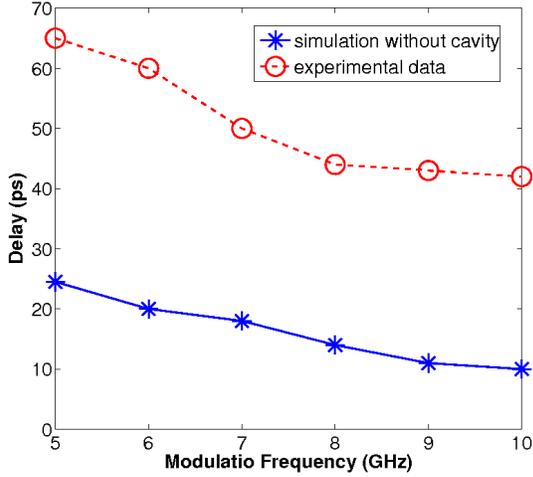}
\caption{Comparison of experimental data and simulation results based on traveling wave CPO model. A big discrepancy is shown without consideration of cavity effect.}
\label{Fig:F2}
\end{figure}

\begin{figure}
\includegraphics[width=8.0cm]{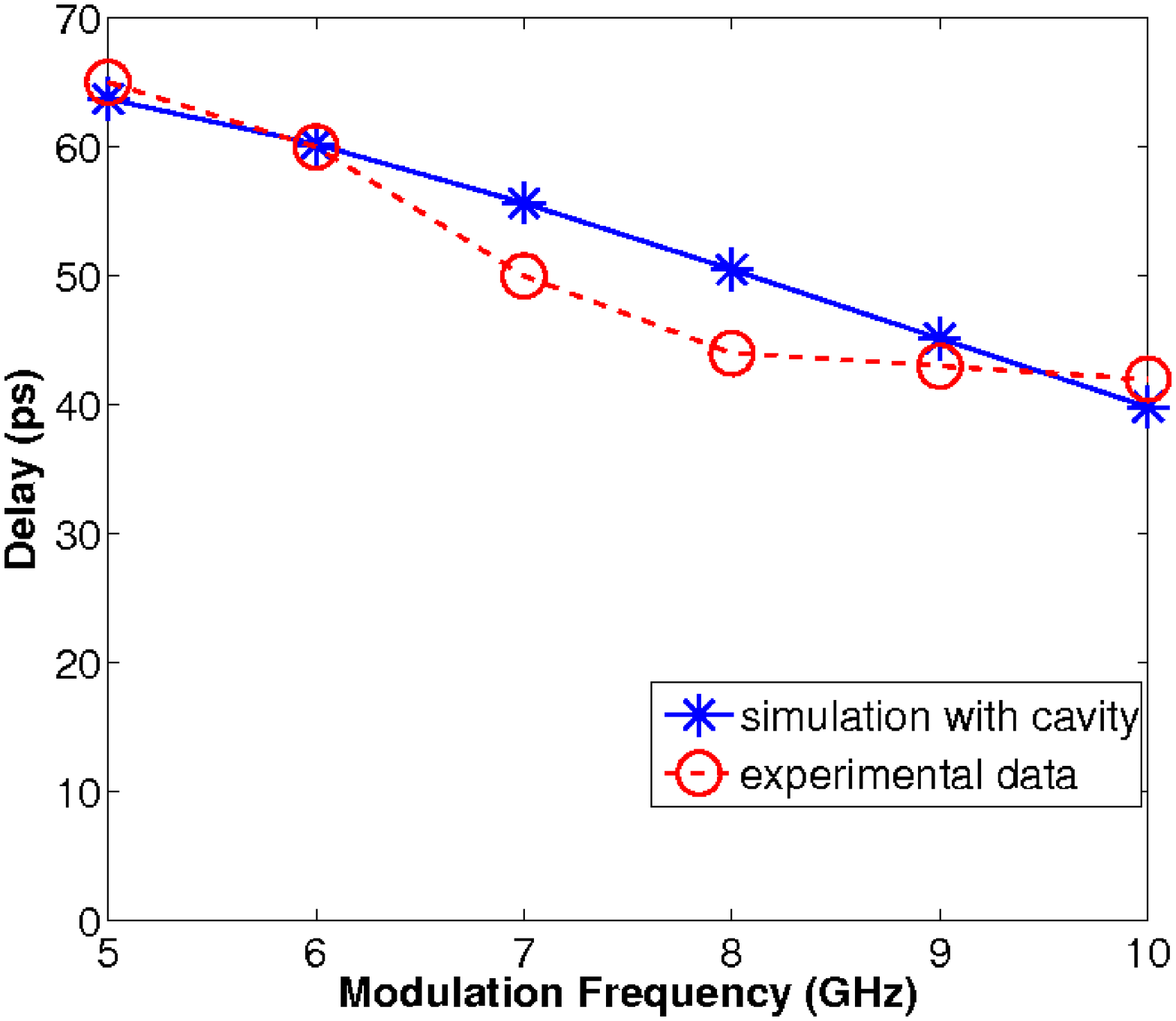}
\includegraphics[width=8.0cm]{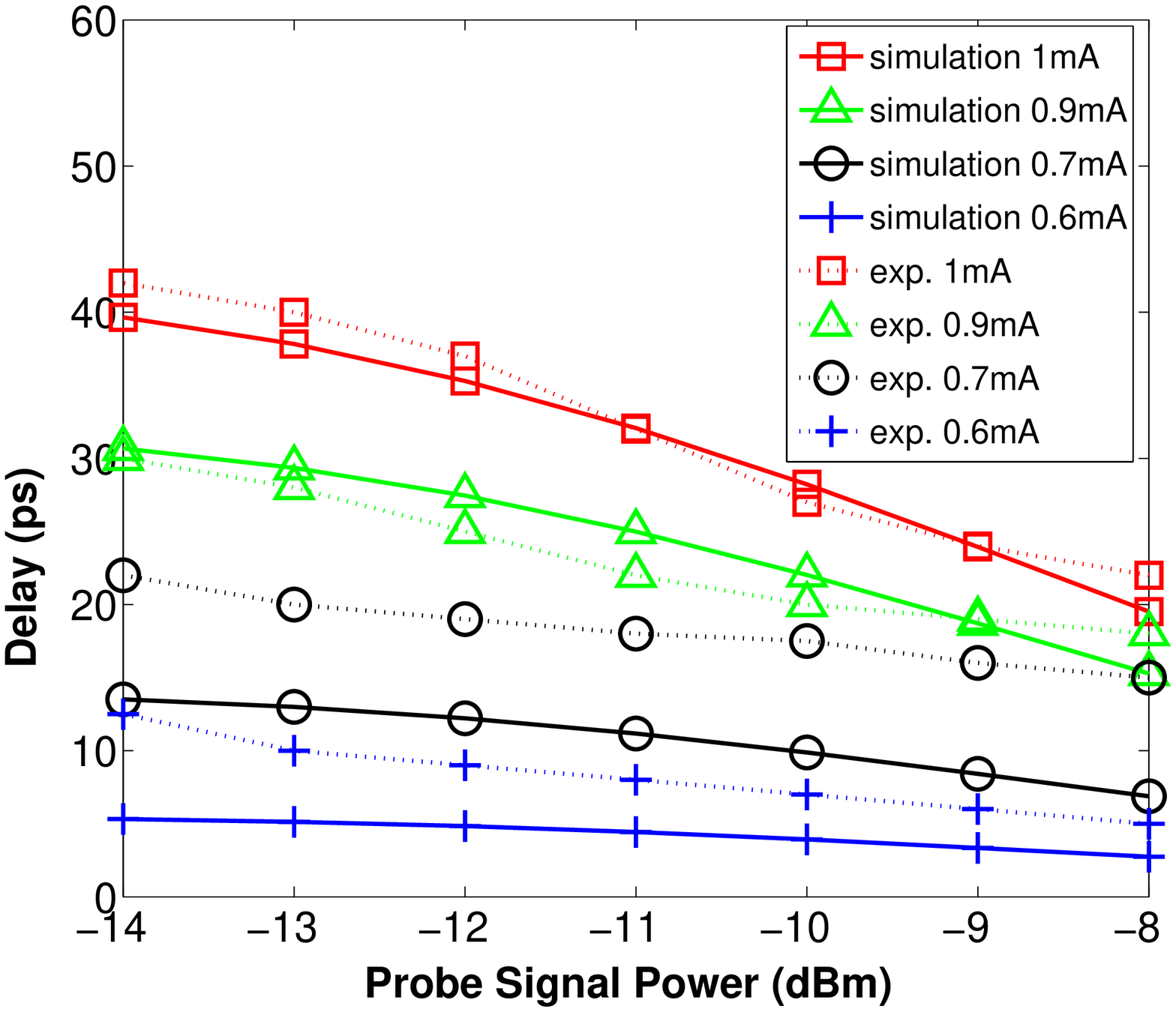}
\caption{(a) Group delay of our VCSELs for different modulation frequency detunings. (b) Group delay is shown as a function of optical power of probe signal beam at different bias currents. Solid lines are simulation results while the dashed lines are experimental data.}
\label{Fig:F3}
\end{figure}

For slow light measurement, a probe signal is generated by a tunable laser and then modulated via an electro-optical modulator.
The signal power is controlled by a variable optical attenuator at the output of the electro-optical modulator.
The wavelength of probe signal is tuned to the resonance of the QD VCSEL cavity at $1.3$ $\mu$m.
An optical circulator is used to couple the probe signal into the QD VCSEL.
The time delay of the reflected probe signal is measured by a digital oscilloscope.
The relationship between the time delays and modulation frequencies of probe signal are shown with the dashed-line in Fig. \ref{Fig:F2}, where the bias current of QD VCSEL and the probe signal power are fixed at $1$ mA and $-14$ dBm, respectively.
The time delay in the QD VCSEL increases as the modulation frequency decreases. 
Moreover, the time delays as functions of bias currents of QD VCSEL and optical power of probe signal are shown with the dashed-lines in Fig. \ref{Fig:F3}(b), where the modulation frequency is fixed at $10$ GHz.
The time delay increases as the signal power decreases. 
The experimental details have been reported in our previous works \cite{PCP-long}.
In this experiment, the threshold current is $0.7$ mA and the thickness of the cavity is estimated as short as about $1.13$ $\mu$m.
\section{MODELING AND SIMULATION RESULTS}
To model the population oscillation in the semiconductor lasers, our theoretical starting point is based on the carrier undulation induced by the frequency beating between two optical waves \cite{SLC-apl06, Agrawal}.
The probe signal experiences gain and refractive index changes by the pump wave through the carrier index and gain grating. The dynamics of the carrier density, $N$, at an injected current, $I$, can be derived from the carrier rate equation, i.e.
\begin{eqnarray}
\frac{d}{d\,t} N = \frac{I}{q V} - \frac{N}{\tau_s} - \frac{g(N)}{\hbar\omega} |E|^2 + D \nabla^2 N(t,z),
\label{Eq1}
\end{eqnarray} 
where $q$ is the unit electron charge, $V$ is the active region volume, $g(N)$ is the model gain, $\tau_s$ is the carrier lifetime, and $D$ is the diffusion coefficient.
$E$ is the field amplitude of total incident waves, including the pump wave $E_p exp(-i\omega_p t)$ and probe signal $E_s exp(-i\omega_s t)$, i.e.
\[
|E|^2 \approx |E_p|^2 + |E_s|^2 + E_p E_s^\ast exp(-i\Omega) t + c.c.
\]
with the the detuning modulation frequency $\Omega$.
A linear model gain is assumed in the model for that our VCSEL is operated not far from the threshold condition,
\[
g(N) = \alpha (N-N_0),
\]
with $\alpha$ is the gain coefficient and $N_0$ is the transparent carrier density.
Next we assume that the carrier density can be described by a dc term and modulated at the same detuning frequency with a small ac terms, 
\[
N \approx \bar{N} + [\Delta N exp(-i\Omega) t + c.c.],
\]
where $\bar{N}$ is the static carrier density and $\Delta n$ is the amplitude of the carrier population oscillation.
The index and gain changes of the probe signal beam can be derived from Eq. (\ref{Eq1}), then one can calculate the corresponding optical group delays caused by the population oscillation effect, with the definition
\begin{eqnarray*}
&&\Delta n_g = \Delta n + \omega \frac{d\, \Delta n}{d\, \omega},\\
&&\tau_{delay} = \frac{L}{c} \Delta n_g,
\end{eqnarray*}
where $L$ is the length of the media and $c$ is the speed of light in free space.
With Eq.(\ref{Eq1}), we assume that the probe signal is much weaker than the pump wave, and obtain the index change  of the probe beam by
\begin{eqnarray}
\Delta n = \gamma g(\bar{N}) \frac{c}{2\omega}[1-\frac{P_0(1+P_0-\frac{\Omega t_s}{\gamma})}{(1+P_0)^2+(\Omega t_s)^2}],
\label{Eqn}
\end{eqnarray}
where $\gamma$ is the line-width enhancement factor, and $P_0 \equiv 
\frac{P}{P_s}$ is the normalized pump power with respect to the saturation power $P_s \equiv \frac{\hbar \omega}{\alpha t_s}$.

Fig. \ref{Fig:F2} shows the comparison of the experimental data of slow lights in our VCSELs with a common adopted CPO model for SOA based on Eq. (\ref{Eqn}).
We use following parameters in the simulations.
The overlap factor is assumed to be $1$, the linewidth enhancement factor $\gamma = 0.5$, the diffusion coefficient $D = 0.8$ cm$^2$/s, and the effective carrier lifetime $t_s = 5$ ns, the transparent carrier density $N_0= 1\times\times 10^{18}$ cm$^{-3}$.
The active region of our VCSEL is approximated by $1$ nm in length, $1$ nm in width, and with thickness of $1.13$ $\mu$m.
We also assume that the gain coefficient is $ \alpha = 2\times 10^{-16}$ cm$^2$ and use $n= 3.2$ as the refractive index.
Experimental operation of the VCSEL shows that the threshold current is $I_{th}= 0.7$ mA, small signal is $g_0=5.53\times 10^6$ (1/m), and the saturation power is $P_s = 1.45\times 10^{-8}$ W.
It can be seen clearly that there is a big discrepancy between the experimental data and the simulations based on the traveling wave CPO model.
Even though the CPO model can predict the tendency of slow light effect for different modulation detunings, the active region of our VCSEL is too shorter to provide enough gain to induce large delays.

Since the main difference between SOA and VCSEL devices is the cavity effect.
In addition to the carrier rate equation in Eq. (\ref{Eq1}), we simplify the DBR cavity in the VCSELs by an effective Fabry-Perot filter with the response of the cavity gain described by \cite{Laurand06},
\begin{eqnarray}
G_r = \frac{(\sqrt{R_t}-\sqrt{R_b} g_s)^2+4 \sqrt{R_t}\sqrt{R_b} g_s \sin^2 \phi}{(1-\sqrt{R_t R_b} g_s)^2 +4 \sqrt{R_t}\sqrt{R_b} g_s \sin^2 \phi},
\label{Eq2}
\end{eqnarray}
where $R_t$ is the top mirror reflectance, $R_b$ is the bottom mirror reflectance, $g_s$ is the single-pass gain, and $\phi_s$  is the single-pass phase detuning.

With the same parameter listed above, the simulation results of the group delays with comparisons to the experimental data for different modulation frequency detuning are shown in Fig. \ref{Fig:F3} (a).
Here the reflectances of top and bottom mirrors are assumed to $R_t=0.997$ and $R_b=0.99$, respectively.
We can see that by including the cavity effect, not only the tendency but also the values of optical delays for different  modulation frequencies at fixed probe signal power and bias current are both in agreement with the real experimental data.
Moreover, the simulation results of the optical delay for different signal powers at different bias currents can fit the experimental observation well without adjusting any parameters, as shown in Fig. \ref{Fig:F3} (b).
The most important signature of our modeling is that the length of the active region is only $1.13$ $\mu$m.
Without including the resonance effect through the cavity in our modeling, there is no possibility to have large group delays up to $65$ ps in such a short semiconductor device.

\section{Discussion and Conclusion}
Based on a simple two-wave model and carrier rate equation, we have a consistent group delay behaviors in VCSELs as the experimental data.
We use population pulsation modal for SOA with additional introduction of cavity effect by applying a Fabry-Perot filter in the theory.
The simulation results of our proposed model agree well with experimental data for different operations of signal power, bias current and modulation frequency detuning with reasonable parameters.
We also compare the simulation differences between coherent population oscillation model with and without cavity effect.
Based on the experimental parameters of quantum dot VCSEL structures, we show that it is possible to have $65$ ps optical group delay within a compact active region as short as $1.13$ $\mu$m.

It is well known that an effective Fabry-Perot filter is no enough to describe the DBR cavity in VCSELs.
Moreover, instead of the two traveling waves used in the population oscillation a standing cavity wave model should be adopted for VCSELs.
And the significant difference between quantum well and quantum dot materials should be classified too. 
A complicated model is under invested for a deep understanding of slow lights in QD VCSELs.
But as a first step, we show that such a simple model can be used for such a compact optical slow light device in the room temperature.
We expect more and more applications based on VCSELs for the applications on the light information storage as well as optics buffer to be happened in the near future.

Authors are indebted to Prof. Shun Lien Chuang for useful discussions.
C.-S. Chou and R.-K. Lee are supported by the National Science Council of Taiwan with contrast NSC 95-2112-M-007-058-MY3 and NSC 95-2120-M-001-006.

\end{document}